\documentclass[twocolumn]{article}

\usepackage{titling}
\usepackage{blindtext} 

\usepackage[sc]{mathpazo} 
\usepackage[T1]{fontenc} 
\usepackage{microtype} 

\usepackage[english]{babel} 

\usepackage[hmarginratio=1:1,top=32mm,columnsep=20pt]{geometry} 
\usepackage[hang, small,labelfont=bf,up,textfont=it,up]{caption} 
\usepackage{booktabs} 

\usepackage{lettrine} 
\usepackage{csquotes}
\usepackage{amsmath}
\usepackage{mathrsfs}
\usepackage{graphicx}
\usepackage{mathtools}

\usepackage{enumitem} 
\setlist[itemize]{noitemsep} 

\usepackage{abstract} 

\usepackage{titlesec} 
\renewcommand\thesection{\Roman{section}} 
\renewcommand\thesubsection{\roman{subsection}} 
\titleformat{\section}[block]{\large\scshape\centering}{\thesection.}{1em}{} 
\titleformat{\subsection}[block]{\large}{\thesubsection.}{1em}{} 

\usepackage{fancyhdr} 
\usepackage{titling} 

\usepackage{hyperref} 

\usepackage{xcolor}

\usepackage{calc,ifthen,setspace,amsmath,amsthm,amscd,amssymb}

\usepackage{csquotes}

\usepackage[ruled,vlined]{algorithm2e}
\usepackage[noend]{algpseudocode}

\newcommand{\R}{\mathbb R}

\newcommand{\e}{\varepsilon}

\newcommand{\calk}{\mathcal{K}}

\newcommand{\calp}{\mathcal{P}}

\newcommand{\calz}{\mathcal{Z}}

\newtheoremstyle%
{custom}%
{}
{}
{}
{}
{}
{.}
{ }
{\thmname{}
\thmnumber{}%
\thmnote{\bfseries #3}}%

\newtheoremstyle%
{Theorem}%
{}%
{}%
{\itshape}%
{}%
{}%
{.}%
{ }%
{\thmname{\bfseries #1}%
\thmnumber{\;\bfseries #2}%
\thmnote{\;(\bfseries #3)}}%

\theoremstyle{Theorem}

\theoremstyle{definition}
\newtheorem{definition}{Definition}[section]

\theoremstyle{remark}

\theoremstyle{custom}

\pretitle{\begin{center}\Huge\bfseries} 
\posttitle{\end{center}} 
\title{Location Privacy in Conservation} 
\author{%
\textsc{Hayyu Imanda} \\[1ex] 
\normalsize University of Oxford \\ 
\normalsize \href{mailto:hayyu.imanda@cybersecurity.ox.ac.uk}{hayyu.imanda@cybersecurity.ox.ac.uk} 
\and 
\textsc{Joss Wright}\\[1ex] 
\normalsize University of Oxford \\ 
\normalsize \href{mailto:joss.wright@oii.ox.ac.uk}{joss.wright@oii.ox.ac.uk} 
}
\date{} 
\renewcommand{\maketitlehookd}{%
\noindent \begin{abstract}

The growing public nature of academic journals along with current best practices of sharing primary data for scientific research are profoundly valuable for the understanding of a species and their conservation efforts. On the other hand, public spatial data on endangered species may be easily abused by wildlife criminals. In this paper, we discuss how geo-indistinguishability, a formal notion of privacy for location-based systems, can be used to add noise to published spatial data whilst allowing quantification of such tradeoff.

\end{abstract} 
}

\begin{document}

\maketitle



\section{Introduction}

A 2019 report from the Intergovernmental Science-Policy Platform on Biodiversity and Ecosystem Services (IPBES) stated that one million plant and animal species are now threatened with extinction \cite{IPBES}. This decline is unprecedented, and is a direct result of human activity. More than ever before, the protection and conservation of these species demand more attention. Among many others, scientists and conservationists have a challenge in responding to this decline.

Looking back in 1999, the gecko \emph{Goniurosaurus luii} from southeastern China was first described by scientists; immediately after they became rarities in the international pet trade, reaching the price of \$1500 to \$2000 per individual. By early 2000s, overexploitation caused extirpation of the local population of \emph{G. luii} \cite{Stuart}. In response to such fact, when an article about the discovery of two new species of \emph{Goniurosaurus} was published in 2015, the authors stated the following:
\blockquote{``Due to the popularity of this genus as novelty pets, and recurring cases of scientific descriptions driving herpetofauna to near-extinction by commercial collectors, we do not disclose the collecting localities of these restricted-range species in this publication. However, such information has been presented to relevant government agencies, and is available upon request by fellow scientists.'' \cite{Goniurosaurus}}

In January 2019, Wiley Publishers announced that the peer-review journal Diversity and Distributions \cite{DiversityDistributions} will join the Wiley Open Access portfolio, resulting in academic articles being free to read for all. This joined many other journals with open-access articles, including Global Ecology and Conservation \cite{GEaC}. In September 2018, cOAlitionS \cite{coalitions} was launched by research funding organisations, with the support of the European Commission and the European Research Council, whose main principle states that ``all scholarly publications on the results from research funded by public or private grants provided by national, regional, and international research councils and funding bodies, must be published in Open Access Journals, on Open Access Platforms, or made immediately available through Open Access Repositories without embargo.'' Similarly, the International Science Council states that openness should be the default for publicly funded research \cite{OpenData}.

The growing public nature of academic publication has caused scientists to debate on their past practices. Reflecting on the \emph{Goniurosaurus} case, Lindenmayer and Scheele \cite{DoNotPublish} stated that unrestricted access to location information for rare, endangered, or newly described species information is facilitating a surge in wildlife poaching. Henceforth, they suggested withholding information for species with high economic value. This triggered a series of open discussions, with Lowe et al. \cite{PublishOpenlyButResponsibly} responding that open-access journals work to ensure sensitive information are securely published, and such data provides tremendous benefits to biodiversity science; not disclosing such information would risk the loss of knowledge which is needed to protect these endangered species. In turn, more responses to such have surfaced \cite{PublishOpenlyButResponsiblyResponse}.

Cooke et al. \cite{Cooke} highlighted that electronic tagging and tracking animals, whilst providing insights to the ecology of wild animals, creates certain issues. Given results are often available without any form of access control, Cooke et al. stated that it is ``necessary to increase the curation, stewardship, and security of electronic tagging information, including tag codes, coding schemes, and receiver-station locations and share data in forms that do not facilitate abuse.'' 

Sharing raw data is fundamental to the advancement of science, due to reproducibility and repeatability practices in the field. In the field of biodiversity, it has been discussed tremendously, with some scientists consider that such practice is a simple ethical principle, while others resist sharing those data. Throughout a study in understanding the varied responses, security and data abuse were not at all mentioned, let alond discussed \cite{Huang2012}. 

It is clear that tradeoffs need to be made in published data and information which may harm conservation of a particular species, and that scientists have not been regularly exposed to the potential security threats that open data might bring. In this project, we propose geo-indistinguishability, a privacy mechanism introduced by Andr\'es et al \cite{geo1} as a tool to help scientists and conservationists maintain location privacy for sensitive spatial data in species conservation, which allows the risk of publishing data to be quantified. This would be done by only publishing noisy data produced by the algorithm, instead of the original data.

In Section \ref{section-maths}, we will briefly discuss the mathematics behind the mechanism. As a case study, in Section \ref{section-casestudy} we will be discussing the critically endangered hawksbill sea turtle, and implement the mechanism on existing data available in the academic literature. We will discuss our results in Section \ref{section-discussion}.


\section{Mathematical Background} \label{section-maths}

Dwork \cite{DworkDP} introduced the notion of \emph{differential privacy}, a mathematical definition which quantifies an individual's risk of participating in a survey. Intuitively, differential privacy ensures that the difference between query results when one individual is removed from the dataset, is multiplicatively not more than a factor parameterised by $\e$. Formally, 

\begin{definition} \label{def-DP}
A randomised algorithm $\mathcal{K}$ gives $\e$-\emph{differential privacy} if for all datasets $D_1$ and $D_2$ differing by at most one element, and all $S \subseteq \text{Range}(\mathcal{K})$,
\[
\Pr[\calk(D_1) \in S] \leq \exp(\e) \cdot \Pr[\calk (D_2) \in S]
\]

One of the attractive properties of differential privacy is that it makes no assumptions about power of (potential) adversary. Similarly, it is independent from auxiliary information that an adversary might have. To achieve differential privacy, one can use the Laplace mechanism, which draws random variables from the Laplacian distribution \cite{DPbook} -- this is a symmetric version of the exponential distribution, and is more focused than the widely-used Gaussian distribution.

\end{definition}

\subsection{Geo-Indistinguishability}

The term \emph{geo-indistinguishability} was introduced by Andr\'es et al. \cite{geo1}, in response to the rise of location-based systems, which include navigation applications, location-based social media, and data mining algorithms. Similar to differential privacy, this is a formal mathematical definition.

The privacy parameter for geo-indistinguishability depends on two factors $\ell$ and $r$. Intuitively, geo-indistinguishability for $(\ell, r)$, or equivalently, to have \emph{$\ell$-privacy within $r$}, means that the user is indistinguishable from any other point within radius $r$ of the actual location, with certainty level $\ell$. That is, $\ell$ would be the distance between the corresponding distributions of the two points separated by at most $r$ distance, so it is sufficient to set a privacy parameter $\e = \ell/r$, and formally define the following:

\begin{definition} \cite{geo2}
Let $\calk$ be a mechanism, with $\calk(x)$ the distribution produced by $x$. Then $\calk$ satisfies $\e$-\emph{geo-indistinguishability} if and only if for all points $x, x' \in \R^2$ with $x \neq x'$,
\[
d_\calp(\calk(x), \calk(x')) \leq \e d_2(x,x'),
\]
where $d_\calp(\cdot, \cdot)$ is the distance between two distributions, and $d_2(\cdot, \cdot)$ the Euclidean metric.
\end{definition}

Note that the definition of $\ell$-privacy within $r$ is equivalent to $\ell'$-privacy within $r'$, where $\ell/r = \ell'/r'$. Equivalently, geo-indistinguishability can be defined as the following: let $\calz$ be a set of points from a mechanism $\calk$ available to an adversary. Then,

\begin{definition} \cite{geo1} \label{def-geo2}
A mechanism $\calk$ satisfies $\e$-geo-indistinguishability if and only if for all observations $S \subseteq \calz$, 
\[
\Pr[S | x] \leq \exp(\e r)\Pr[S | x'],
\]
for all $r \geq 0$, for all $x, x'$ such that $d_2(x, x') \leq r$.
\end{definition}

As we see, Definition \ref{def-geo2} very much corresponds to Definition \ref{def-DP} of differential privacy by taking into account an arbitrary metric between databases. Hence, geo-indistinguishability can be seen as a generalisation of differential privacy, using Euclidean distance.

To achieve geo-indistinguishability, the authors looked into the Laplacian distribution, as how most of differential privacy mechanisms are designed. However, given that we are looking at two-dimensional spatial data, the authors defined a planar Laplacian distribution, a two-dimensional natural extension of the Laplacian distribution. Given privacy parameter $\e \in \R_{>0}$ and true location $x_0 \in \R^2$, the  probability density function of the planar Laplacian distribution on a point $x \in \R^2$ with $x \neq x_0$ is
\[
D_\e(x_0)(x) = \frac{\e^2}{2\pi}\exp(\e d_2(x_0,x)).
\]
Choosing a point generated randomly based on the distribution above is called the planar Laplacian mechanism and, indeed, can be shown to satisfy geo-indistinguishability \cite{geo1}.

The planar Laplace distribution can be translated to a system of polar coordinates, whereby a point $x$ can be described with parameters $r = r(x_0)$ and $\theta = \theta(x_0)$ from $x_0$. The probability density function of the polar Laplace centred at $x_0$ is
\[
D_{\e}(r, \theta) = \frac{\e^2}{2\pi}r\exp(-\e r),
\]
and in fact, this brings along a convenient property; let $R, \Theta$ be the random variables representing the radius and the angle respectively. In fact, the two variables are independent:
\[
D_{\e}(r, \theta) = D_{\e, R}(r) \cdot D_{\e,\Theta}(\theta). 
\]
Hence, to efficiently draw a point from the distribution, it is sufficient to draw $r$ and $\theta$ separately. Since $D_{\e, \Theta}(\theta) = \frac{1}{2\pi}$ is constant, $\theta$ can be drawn from a uniform distribution on the interval $[0,2\pi)$. To draw $r$, as shown in \cite{geo2}, it is sufficient to generate a random number $z$ from the uniform distribution on the interval $[0,1)$ and set 
\[
r = -\frac{1}{\e}\left(W_{-1}\left(\frac{z-1}{e}\right) + 1\right),
\]
where $W_{-1}$ is -1 branch of the Lambert W function. We will use these properties in our implementation.

\section{Case Study: \emph{Eretmochelys imbricata}} \label{section-casestudy}

The hawksbill sea turtle (\emph{Eretmochelys imbricata}) is found throughout the coral reefs and hard-bottom habitats in the waters of the tropical and subtropical seas in the Atlantic, Pacific, and Indian oceans. It is critically endangered, with its population decreasing and humans one of their primary threats \cite{IUCN}; hawksbill meat and eggs are collected for food, the oil in their muscles for traditional medicines, and their colourful shells for decorative pieces. International trade on tortoiseshell was banned in 1957 by CITES among its signatory nations, with Japan, known for its \emph{bekko} trade, joining only in 1992. Regardless of such regulations, multiple mass poaching of hawksbill sea turtles are still reported since, including by Chinese vessels operating illegally in Indonesia and Malaysia in 2007 \cite{SWOT}. It is estimated that almost 9 million hawskbill turtles were harvested in networks concentrated in South East Asia in the past 150 years \cite{Millereaav5948}. 

The hawksbill sea turtle is an interesting case to study due to their high poaching rate, but also due to prevalence of spatial information on their movements, especially nesting sites. Some of these nesting sites are described to be within a small geographical area (e.g. on a specific beach on an island). Such information are available in different countries and areas, including the Dominican Republic \cite{DominicanRepublic}, Seychelles \cite{Seychelles}, Barbados \cite{Barbados}, and Persian Gulf \cite{PersianGulf}. In \cite{Arabian},  location maps on complete movements (along their GPS coordinates) of 90 post-nesting tagged turtles in the Arabian region were specified.

\begin{figure}[htb]
\centering
\includegraphics[width=7cm]{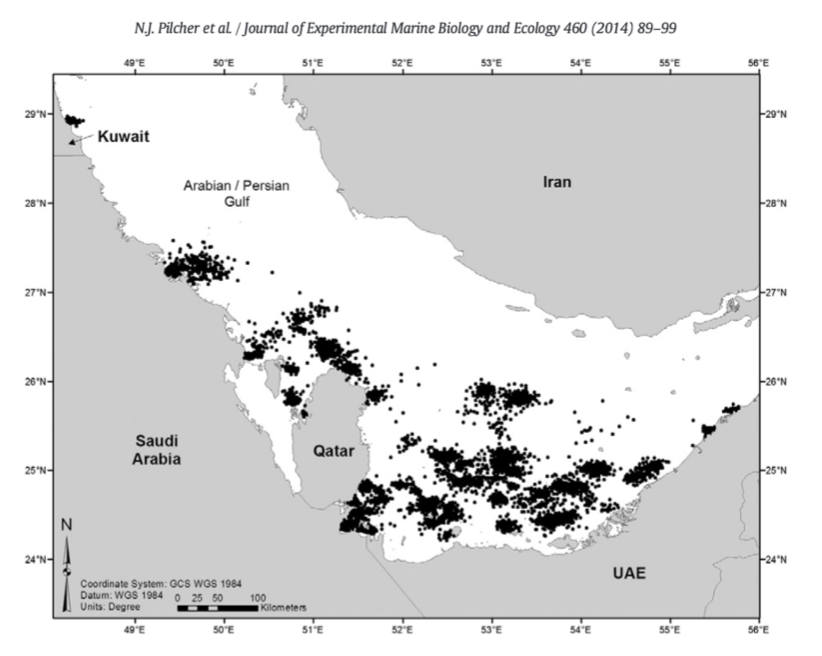}
\caption{Locations of individual hawksbill turtles foraging grounds in the Arabian gulf \cite{Arabian}.}
\end{figure}

In this project, we will focus on two papers containing spatial data of hawksbill turtles: the first is a study of spatiotemporal preferences of the turtles in Melaka, Malaysia \cite{Melaka}, which contains GPS coordinates of beaches studied, as well as the number of nest per beach, which is nonuniform and with a high preference in woody vegetation zones. The paper also summarises its temporal findings -- the turtles in Melaka nest year round with peaks between May and August, with peak time between 10pm and midnight. Such spatial and temporal nesting preferences, to the authors' opinion, would contribute towards the conservation of the endangered species.

Secondly, a study on hawksbill aggregation in Florida was conducted within a 30-km stretch ($26^{\circ}55'$N to $26^{\circ}37'$N) off Palm Beach County, Florida \cite{Florida}. 435 scuba dives were conducted across 44 dive sites, with specific GPS coordinates published, alongside information on hawksbill sightings. This is an interesting case, given oceanic conditions might change the threat model for abuse of spatial information.

\section{Implementation and Results} \label{section-implementation}

A Python script which inputs a set of GPS coordinates $(lat, long)$ and a privacy parameter $\e$ was written, to output a new set of GPS coordinates with added noise, which is $\e$-geo-indistinguishable from the original point.

\begin{algorithm}[h!] \label{algorithm-laplace}
\caption{Geo-Indistinguishability}
\KwIn{$(lat, long), \e$}
\KwOut{$(shiftedlat, shiftedlong)$}
	$\theta \leftarrow [0, \frac{\pi}{2})$ \\
	$ z \leftarrow [0, 1)$ \\
	$ r = -(W_{-1}(\frac{z-1}{e})+1)/\e$ \Comment{\textcolor{gray}{selecting the radius from the distribution}} \\
	$(x_1,y_1) = \verb+degreeToRad+(lat, long)$  \Comment{\textcolor{gray}{translating into polar coordinates}} \\
	$(x_2, y_2) = \verb+addVector+(x_1, y_1, \theta, r)$ \Comment{\textcolor{gray}{adding a vector of length $r$, angle $\theta$ to $(x_1, y_1)$}}\\ 
	\Return $\verb+radToDegree+(x_2, y_2)$ \Comment{\textcolor{gray}{translating back to latitude, longitude}}
\end{algorithm}

Algorithm 1 adds noise taken from the planar Laplace distribution introduced in \cite{geo1} to the original coordinate, and is similar to that used in Location Guard \cite{LocationGuard}. We used \verb+mpmath+ \cite{mpmath}, a Python library which implements the Lambert W function.

Our implementation is available on \url{https://github.com/himanda/geo-indistinguish}.

\subsection{Validation}

\begin{figure*}[htb] \label{fig-laplace}
\centering
\includegraphics[width=12cm]{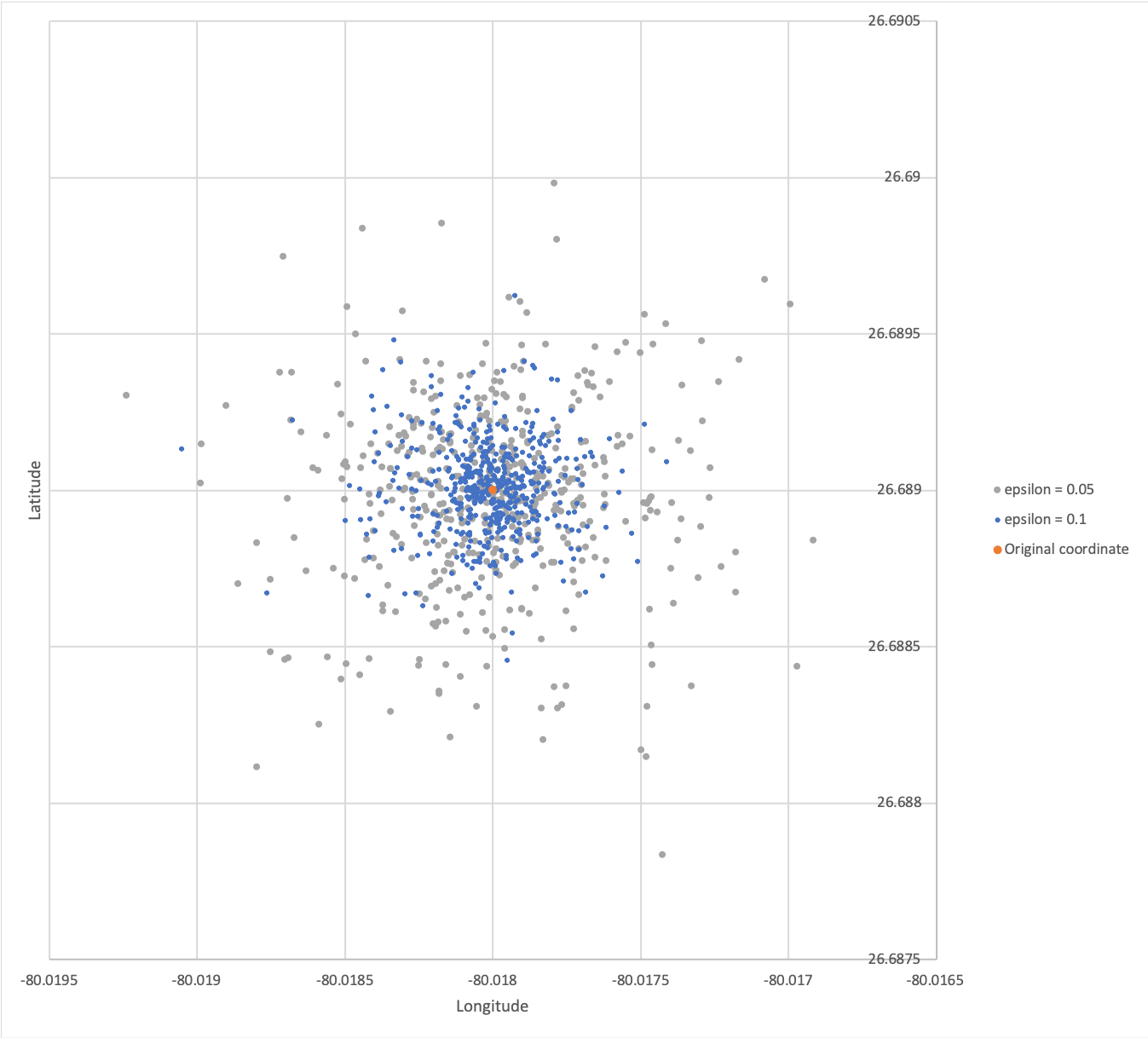}
\caption{Results of Algorithm 1 to Ron's Reef dive site \cite{Florida} showing the distribution with 512 data points for $\e = 0.05$ and $\e = 0.1$.}
\end{figure*}

We ran Algorithm 1 to the following two sets of GPS coordinates:
\begin{enumerate}
\item $(26.689, -80.018)$: Ron's Reef dive site specified in \cite{Florida}, a sighting area for hawksbill turtles.
\item $(2.3161, 102.0704)$: Padang Kemunting beach in Melaka, Malaysia \cite{Melaka}, as a site of hawksbill turtle nests.
\end{enumerate}

To show the planar Laplacian distribution, we have computed 512 coordinate points for each $\e = 0.05$ and $\e = 0.1$, as shown in Figure 2. As we expect, given that $\e$ is inversely proportional to the radius, smaller $\e$ would lead to data which is more sparse, and is of a larger distance to the original point. Table 1 shows the average distance of the noisy point from the original point, from 512 runs of Algorithm 1 to the coordinate $(26.689, -80.018)$.

\begin{table}[htb] \label{table-distance}
\centering
\begin{tabular}{l | c}
$\e$ & Average Distance (m)\\
\hline
0.5 & 3.93\\
0.2 & 10.15 \\
0.1 & 19.48 \\
0.05 & 37.87 \\
0.02 & 98.40 \\
0.01 & 196.74 \\
0.005 & 391.96 \\
0.002 & 942.17 \\
0.001 & 1935.13
\end{tabular}
\caption{Average distance of 512 shifted coordinates pairs from the original point.}
\end{table}

In differential privacy, the choice of privacy parameter $\e$ is not necessarily a technical problem, but arguably a social one. Similarly, in this scenario, choosing a privacy parameter requires understanding of threat model as well as local topographical features. For example, the choice of $\e$ for the Ron's Reef dive site could take into account that it is a dive site; though turtles do swim up to the surface at times, it is a fair assumption that access to a dive site 2 kilometres away from shore causes limitations for attackers. Further, in Florida, the waters are more likely to be patrolled, as well as, due to currents and technical limitations, a small change in distance away from a dive site might result in a completely different underwater sightings. Hence, in this case, the privacy parameter $\e$ could be selected to have a low privacy level $\ell$ and a low radius $r$.

Other topographical features also need to be considered. For example, when the beach is on a small island surrounded by water, then it makes no sense to publish coordinates out in the ocean, so the privacy parameter needs to be adjusted accordingly.

\begin{figure}[htb] \label{figure-nest}
\centering
\includegraphics[width=7cm]{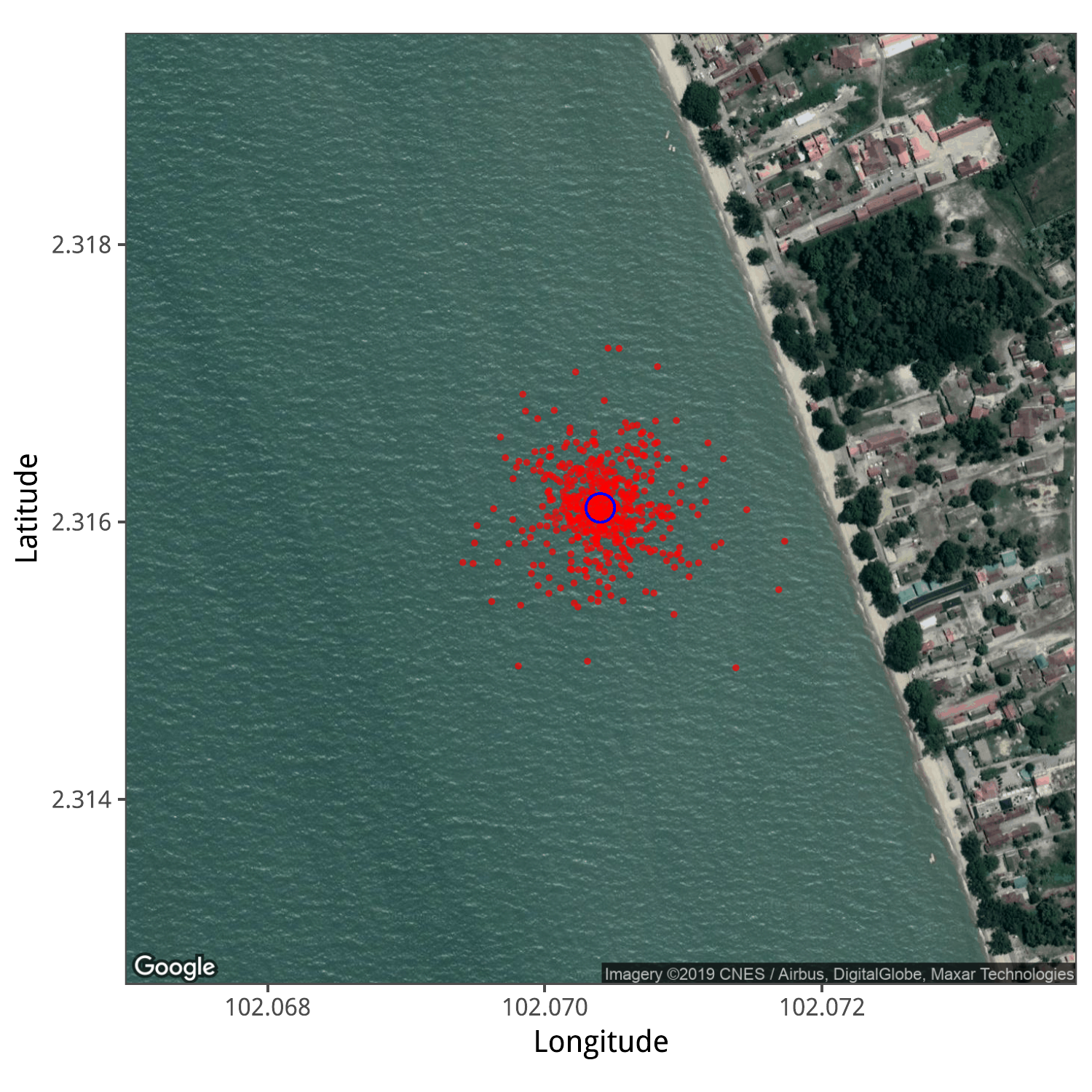}
\caption{Result of Algorithm 1 to Padang Kemunting beach with 512 data points for $\e = 0.05$.}
\end{figure}

On the other hand, when individual nest site coordinates are recorded, this provides an interesting case. Though our case study provides GPS coordinates of the beach (along with the number of nests) instead of the individual nest themselves, we assume that we still want to protect those coordinates. A potential issue with such geography is that nesting sites are within a strip of a beach, which resides between human habitation and water, and hence, the planar Laplace mechanism can output a point in either one. This can of course be mitigated by starting the algorithm over again until a beach coordinate is found, but of course that can lead to an attack by an adversary.

\section{Discussion and Further Work} \label{section-discussion}

Firstly, we are not at all claiming that using publicly available data is a common practice amongst poachers at the moment, but we are saying that it is possible for them to do so. If we are still modelling the threats in a traditional way, we will not be able to face threats that will arrive in the near future. Indeed, when poachers in India obtained access to radio tracking collar data to track and capture endangered tigers, many news agencies has been keen to add the prefix to `cyber-poaching' \cite{cyberpoaching}. 

Originally, we looked at citizen science projects such as Wildbook \cite{wildbook}, which allow public submissions of pictures of wild animals along with spatiotemporal data, which can then be personally identified using machine learning. This project is open-source, and though can have tremendous benefits for researchers, it can also act as tool for poachers to track endangered animals for capture. It would be interesting to see whether differential privacy can be used in its database queries, whilst quantifying the risk of an individual information being shared. Similarly, geo-indistinguishability can also be implemented to the geotag data of the animal, implementing Algorithm 1 in the system.

Due to time concerns, tradeoff analysis have not been properly quantified in our case study. Ideally, we would like to analyse thoroughly the $\e$ parameters in those specific topographies, levelled with the scientific aim of the paper. That is, we would like to provide a range for $\e$ in which the same conclusions can be drawn from the paper, and hence provide an upper bound for $\e$ in which the data becomes useless. The choice of $\e$ parameter requires collaboration across disciplines.

At the moment, our algorithm is implemented on static data, but it of course can be implemented to nonstatic objects as well, e.g. animal tags, with a tradeoff of efficiency and hardware limitations. This of course brings a further debate in other areas -- park rangers use animal tags to ensure tourists and visitors have a high chance of seeing animals, but such certainty will vanish when geo-indistinguishability is introduced. 

In the case of topographical limitations (e.g. a beach in our case study), we might be interested in constructing an attack which takes into consideration the surrounding topography of the published location. Alternatively, one can find a new mechanism which truncates unwanted parts in its distribution, while maintaining the definition of geo-indistinguishability only in the applicable areas.

There is also an issue of interdependence that needs to be studied further. When multiple locations are published in close proximity to one another, could we construct an attacks which can efficiently guess the actual locations? Further, can we use the symmetricity of the planar Laplacian distribution to help construct an attack?

Lastly, an ideal goal of the studies to follow this project is to have a tool for researchers, where they can simply plug in spatial information, and input the privacy parameter with topographical considerations in mind, and outputs a new coordinate which is geo-indistinguishable from the original point, which they can in turn publicly share. Of course, they still have the raw data which they may share with other entities, or with an established access control mechanism to mitigate the issues of repeatability and reproducibility.


 

\section{conclusion}

We stress the importance of security considerations in publishing conservation spatial data. We further introduce a novel approach to location privacy in conservation by the use of geo-indistinguishability, a privacy-preserving mechanism generalised from differential privacy, which allows risk quantification for published data.



\end{document}